\documentclass[11pt,twoside]{article}

\usepackage{asp2006}
\usepackage{epsf}
\usepackage{psfig}
\usepackage{lscape}
\usepackage{graphicx}

\def\teff{T$_{\rm{eff}}\,$}
\def\teffs{T$_{\rm{eff}}s\,$}

\def\Dwa{$\,$\uppercase\expandafter{\romannumeral5}$\,$}
\def\mic{$\mu$m$\,$}

\def\sless{\lower2pt\hbox{$\buildrel {\scriptstyle <}
   \over {\scriptstyle\sim}$}}
\def\sgreat{\lower2pt\hbox{$\buildrel {\scriptstyle >}
   \over {\scriptstyle\sim}$}}

\begin{document}
\title{The Role of Dust Clouds in the Atmospheres of Brown Dwarfs}   
\author{Adam Burrows}   
\affil{Department of Astrophysical Sciences, Princeton University, 
Princeton, NJ 08544; burrows@astro.princeton.edu}    

\begin{abstract} 
  The new spectroscopic classes, L and T, are defined by the
role of dust clouds in their atmospheres, the former by their presence
and the latter by their removal and near absence. Moreover, the
M to L and L to T transitions are intimately tied to the condensation
and character of silicate and iron grains, and the associated clouds play
pivotal roles in the colors and spectra of such brown dwarfs.
Spanning the effective temperature range from $\sim$2200 K to $\sim$600 K, these
objects are being found in abundance and are a new arena in which
condensation chemistry and the optical properties of grains is
assuming astronomical importance.  In this short paper, I 
summarize the role played by such refractories in determining the
properties of these ``stars" and the complexities of their theoretical
treatment.
\end{abstract}


\section{The Importance of Dust Clouds in Cool Atmospheres}
\label{importance}

\begin{figure}
   \includegraphics[width=3.0in,height=2.0in,angle=-90]{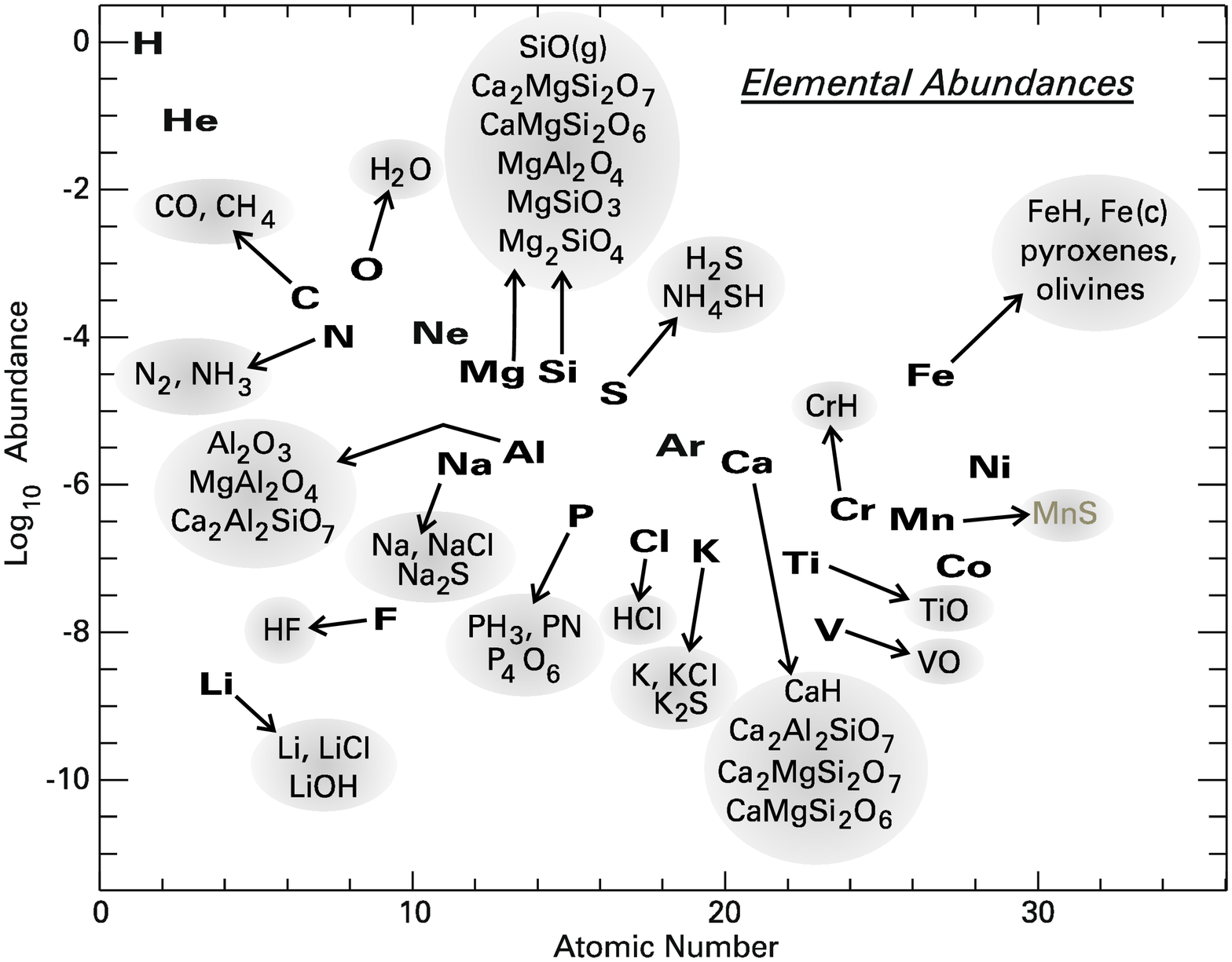}
\vskip5.0in
\caption{This figure depicts the trend of elemental abundance
with atomic weight and identifies the dominant chemical forms
of each in the context of brown dwarf atmospheres.\label{fig:1}}
\end{figure}

\begin{figure}
   \includegraphics[width=5in, angle=0]{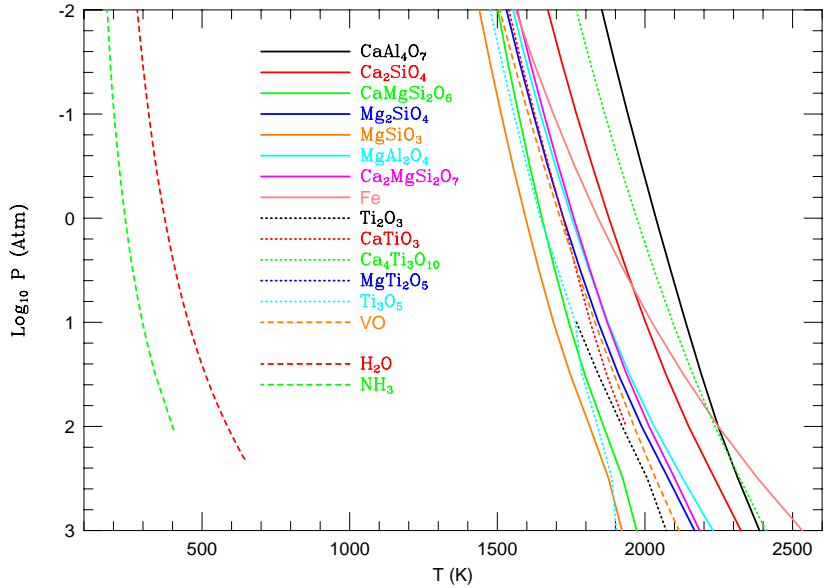}
\caption{Solar-metallicity condensation curves (temperature
in Kelvin versus pressure in atmospheres) for many of the most important refractory species
thought to appear in the atmospheres of brown dwarfs and L dwarfs.  The most refractory
compound is the calcium aluminate grossite  
(CaAl$_4$O$_7$ $\equiv$ CaO + 2(Al$_2$O$_3$)).  Corundum (Al$_2$O$_3$),
as such, does not generally form.  CaMgSi$_2$O$_6$ is diopside,
Mg$_2$SiO$_4$ is forsterite, MgSiO$_3$ is enstatite,
MgAl$_2$O$_4$ is spinel, and Ca$_2$MgSi$_2$O$_7$ is akermanite.
The dotted curves correspond to the refractory titanium compounds.
Liquid Fe is the solid curve at a slightly
shallower slope than those for the calcium/aluminum/magnesium condensates (solid).
Included for comparison are the condensation curves for water (H$_2$O) and ammonia
(NH$_3$).  Notice how the condensation curves
of the refractory compounds densely inhabit a narrow range of T/P space and that there is
a noticably wide gap between this refractory band and water.
See text for a discussion of the salient features of this figure.
[Taken from Burrows, Sudarsky, \& Hubeny 2006]\label{fig:2}}
\end{figure}

As the temperatures of a stellar atmosphere decrease below $\sim$4000 K, 
molecules form and begin to dominate.  Water and carbon monoxide are two of the
first to make their mark, but despite the low elemental abundance of titanium
($\sim$10$^{-7}$) and vanadium, TiO and VO too emerge in the M dwarf range 
as distinctive signatures in the optical.  Figure \ref{fig:1} portrays the elements
in order of abundance, shows the major molecules into which these elements 
partition as the temperature decreases, and suggests which species predominate 
due to their relative abundance. However, as \teff decreases further 
and approaches $\sim$2200 K, many of the more refractory elements
condense out into clouds of dust.  Figure \ref{fig:2} depicts many of the corresponding 
condensation curves at solar metallicity. Titanium begins to form perovskite (CaTiO$_3$)
and higher oxides, followed near $\sim$1700$-$1800 K by vanadium, which first 
forms condensed VO.  Importantly, calcium-aluminum and 
calcium-magnesium silicates (such as akermanite, diopside, hibonite, and grossite) 
form and sequester many refractory elements into grains, whose depletion is 
indirectly manifest by the gradual disappearance of atomic lines of titanium,
calcium, aluminum, and silicon.  More importantly, a haze is formed that thickens
into formidable clouds whose continuum opacity begins to redden the object's
near-infrared spectrum. This reddening signals the appearance of
the new spectroscopic class of L dwarfs. Indeed, the M to L transition is caused by
the formation of dust (and the simultaneous disappearance of TiO and VO).
The entire L dwarf sequence is dominated by the prevalence of dust clouds and 
is defined by red $J-K$ colors.  Since the calcium and aluminum abundances are $\sim$10$\times$
smaller than the magnesium and silicon abundances, trapping the former 
into the most refractory compounds in stoichiometric ratios leaves plenty of the 
latter to form magnesium silicates, such as enstatite (MgSiO$_3$) and forsterite (Mg$_2$SiO$_4$),
at temperatures below $\sim$1850 K.  These species of dust, along with iron droplets
and numerous compounds all along the olivine and pyroxene sequences, constitute
the opaque clouds that determine L dwarf properties.  

\begin{figure}
   \includegraphics[width=5in, angle=0]{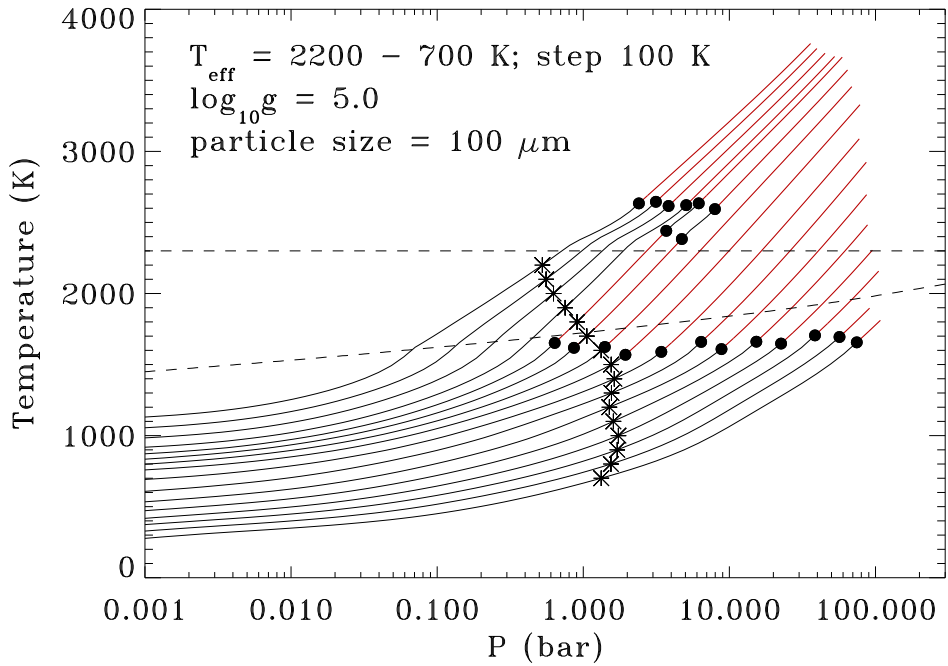}
\caption{Temperature-pressure profiles for a set of models
with constant surface gravity, $\log_{10}g({\rm cm\ s}^{-2}$) = 5.0 and
particle size, 100 microns, for different effective temperatures
ranging from \teff = 2200 K (leftmost curve) to 700 K
(rightmost curve). The positions where the local temperatures are equal
to the effective temperatures (which indicate approximate locations
of the photospheres) are shown as asterisks.
The cloud bases are depicted as dashed lines;
the black dots show the position of the boundaries of the convection
zone(s). Notice the occurence of two distinct convection zones
for \teff  = 1700 K and 1800 K.[Taken from Burrows, Sudarsky, \& Hubeny 2006]\label{fig:3}}
\end{figure}

Since the base of a cloud is found near its condensation line,
which in temperature-pressure space is mostly a function of temperature,
a cloud's geometric thickness may be some fraction of a pressure scale height. As \teff decreases,
though the cloud's optical thickness increases further it is progressively 
more deeply buried.  Figure \ref{fig:3} portrays the positions of the radiative-convective boundaries,
the photospheres, and the realm of the relevant condensation curves for a \teff\ sequence 
of self-consistent atmosphere models.  When \teff reaches $\sim$1000-1200 K, the $\sim$1500$-$2200 K region 
of the atmosphere where refractory clouds reside has been buried so deeply that 
silicates, though present, are of secondary importance in the emergent spectrum.  
At this point, the atmosphere is depleted of refractory metal elements and
emerges into the T dwarf realm.  This is the L to T dwarf 
transition, characterized by such a clearing. Hence, the appearance and disappearance
of dust is of central importance in distinguishing the M, L, and T dwarf spectral classes.
Were it not for dust, the L dwarfs would not exist as a spectroscopic type.

Moreover, the edge of the hydrogen-burning main sequence, in fact what determines a star,
is in the middle of the L dwarf sequence (not at the end of the M dwarf sequence!). Hence,
the opacity of silicate grains plays a central role in determining what is and is not a star.
Currently, we estimate that the solar-metallicity stellar edge is near L4/5, a \teff 
of $\sim$1700 K, a bolometric luminosity of $\sim$6$\times$10$^{-5}$ L$_{\odot}$,
and a mass of $\sim$0.074 M$_{\odot}$ (Burrows et al. 2001), but we really don't know. This is a curious
state of affairs after more than 100 years of astrophysics and is one more indication
of the importance of dust in astronomy.

\section{Complications of Cloud Modeling}
\label{complex}

Unfortunately, to understand the M$\rightarrow$L$\rightarrow$T
sequence and their spectra in detail requires a mastery of not only the chemical
condensation sequences and the consequent elemental depletions (``rainout"; Burrows \& Sharp 1999)
in a gravitational field, but the spatial extent, particle size and shape distributions,
grain optical properties, and meteorology of clouds as well.  These complications do not confront
one who models most other types of ``stars" and make brown dwarf theory rather more challenging.

These challenges have not yet been adequately met. There are many reasons, 
a few of which I now identify. At a given pressure, the refractories included 
in Fig. \ref{fig:2} appear in a narrow range of temperatures. Furthermore, just 
a bit later than the early L dwarfs the optical depths of such clouds are likely 
to be sufficient to trip convection (and the associated updrafts and downdrafts) 
where there are clouds.  Hence, after the early Ls (for which the first condensates 
inhabit a stably-stratified radiative zone) every condensate whose condensation curve
intersects a dwarf's $T/P$ profile will most probably reside in a common convection zone.
It does not make sense to assume that each condensate is a separate, isolated layer. 
Rather, as the \teff\ decreases and the first clouds thicken, 
convection is tripped in the atmosphere. The kinetics of such a soup of growing condensates
in or out of a convective zone is a daunting problem, though 
grain growth in the brown dwarf context has been receiving some attention of late
(Ackerman \& Marley 2001; Cooper et al. 2003; Helling et al. 2001, 2004; 
Woitke \& Helling 2003,2004).  Note that below the cloud base in the inner convective zone
is the ``infinite" reservoir of heavy elements that extends throughout the dwarf and that 
sets the heavy-element abundance boundary condition.  However, the heavy elements
that may have once existed in the upper atmosphere before condensation do not all remain
in the cloud once formed.  How much cloud material does remain in the cloud depends on the dynamics
of the cloud itself.  

Clearly, the equilibrium particle size 
distribution, achieved through the balance of growth processes in the 
convective zone and grain destruction at and below the $T/P$-profile/condensation-curve 
intercept, is very poorly constrained by theory. Furthermore, the optical
constants of heterogeneous grains of indeterminate composition
and layering are not easily derived from first principles.  Qualitatively, it is clear
that the modal particle size of grains in stable radiative zones is smaller ($\sim$0.1$-$5.0 \mic)
than in turbulent convective zones ($\sim$10$-$150 \mic),
but confidence in the current analytic estimates should not be great.
Unfortunately, modeling requires a handle on particle growth and size, composition, optical properties,
and cloud spatial extent, all in the context of a consistent radiative-convective atmosphere model
with, perhaps, non-equilibrium chemistry.

\section{Anomalies in Brown Dwarf Atmospheres}
\label{anomaly}

\begin{figure}
   \includegraphics[width=5in, angle=0]{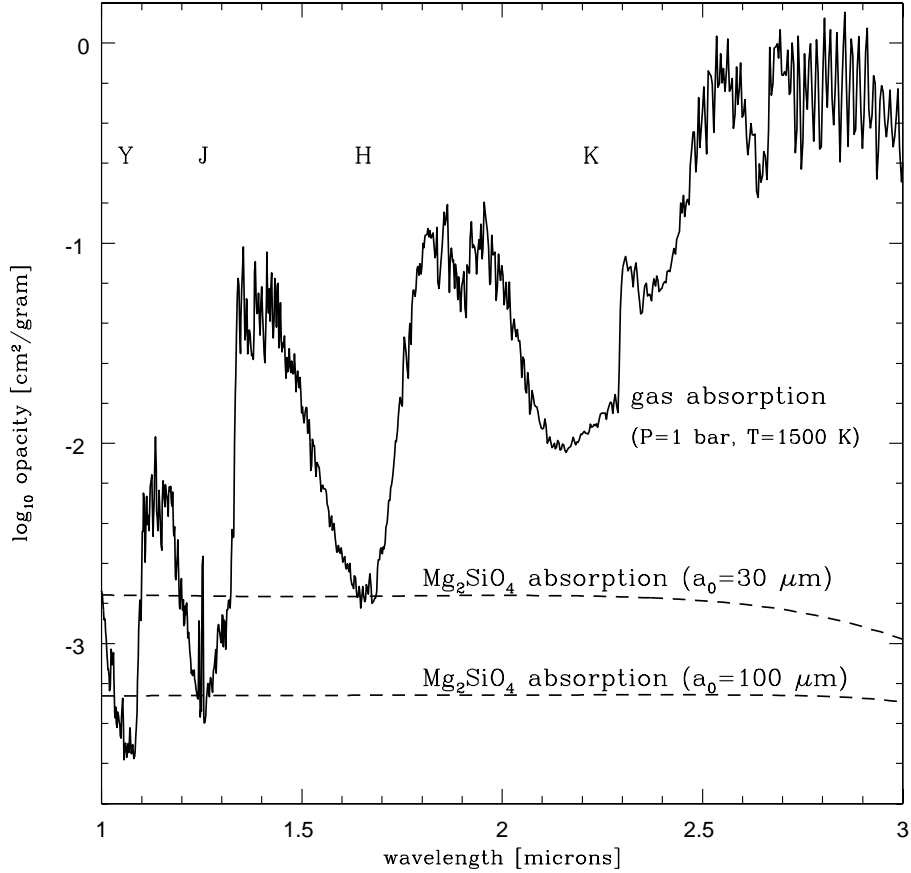}
\caption{Abundance-weighted comparison of forsterite opacity
for 30-\mic and 100-\mic modal particle sizes with that of the
total gas opacity at a pressure of 1 bar and temperature of
1500 K.  In the $Y$/$Z$ and $J$ bands, forsterite can be a dominant
opacity source, depending upon the depth of the cloud layer
in the atmosphere.[Taken from Burrows, Sudarsky, \& Hubeny 2006]\label{fig:4}}
\end{figure}

There are numerous unexplained facts concerning brown dwarf spectra, many of which 
are related to dust physics.  The brightening in the $J$ band (Dahn et al. 2002; Tinney, Burgasser, \&
Kirkpatrick 2003; Vrba et al. 2004) is the most intriguing anomaly . Its explanation must
be the rapid thinning out of the clouds in the spectrum-forming region of the brown dwarf atmosphere
during the L$\rightarrow$T transition.  Figure \ref{fig:4} indicates why on generic grounds one would expect
such a brightening to be in the $J$ and $Y/Z$ ($\sim$1.0$-$1.1 \mic) bands, if anywhere. 
But how the effective opacity of the silicate clouds decreases so quickly with \teff and 
spectroscopic subtype to yield heavy-element depleted T-dwarf atmospheres has yet to be explained.
Current models are not adequate (Tsuji, Nakajima, \& Yanagisawa 2004; 
Allard et al. 2001; Marley et al. 2002; Burrows, Sudarsky, \& Hubeny 2006).
Burgasser et al. (2004) have postulated the break up of the clouds near a \teff of 1200$-$1300 K
and the appearance of holes, whose filling fraction increases across the L$\rightarrow$T transition
until that fraction is unity.  A virtue of this model is the natural explanation
of the apparent resurgence of the FeH features in the early- to mid-T dwarfs
(Burgasser et al. 2004; McLean et al. 2003; Cushing, Rayner, \& Vacca 2005).
The FeH abundances near the photospheres should be waning; holes
could allow us to see more deeply to the higher-temperature regions in which the FeH abundance is
large.  Knapp et al. (2004) suggest an increase in the ``sedimentation efficiency"
of the clouds, with a concommitent rapid increase in the silicate particle size.
Liu \& Leggett (2005), Burgasser et al. (2005), and Burrows, Sudarsky, \& Hubeny (2006)
suggest some role for binarity (``crypto-binarity") at the L to T transition.
The binary fraction of T dwarfs is not negligible and $\sim$0.75
mag (2.5$\times\log_{10}2$) is near the magnitude of the few
excesses measured. 

Burrows, Sudarsky, \& Hubeny (2006) have shown that when calcium-aluminate, silicate, and Fe clouds first form
they do so in the radiative region, that as \teff\ decreases an isolated convective
zone emerges, and that for even lower \teffs\ the two convective zones join.  
Figure \ref{fig:3} depicts this merger and the relative positions of the photospheres
and the radiative-convective boundaries.  However, it remains to be seen 
what happens to the particle sizes and cloud morphology when these regions merge.
Shaw (2003) and Kostinski \& Shaw  (2005) investigate the
dependence of runaway droplet growth and rainout on the presence in convective
clouds of large velocity shears and on intermittency in the turbulence.  Could
the merger of the outer convective cloud with the inner convective zone
lead to regions of such large shears, in which particle growth on the
timescales available is more rapid, and, hence, lead to very large particles?
Could the merger lead to the irreversible partial flushing of cloud material into the interior?
After the joining of the convective zones, is the timescale for grain growth too
long for the convecting feedstock to avoid being dragged into the hot interior
before forming opaque grains?  Is a critical \teff/gravity threshold for rapid
grain coalescence and growth reached, beyond which the average particle is too
large to contribute significant opacity (Liu, Daum, \& McGraw 2005)? Or does the scale
height of the silicate cloud collapse at some \teff\ threshold? The answers to
these questions require a multi-dimensional approach both to grain kinetics and growth
and to convective cloud structures and motions, all properly coupled.

\section{Rainout and the Importance of Alkali Metals}
\label{rainout}

One of the curiosities of brown dwarf spectra is the prevalence 
and importance in the optical and near infrared of just
two doublets, the sodium D lines centered at 0.589 \mic and the corresponding 
potassium resonance lines at $\sim$0.77 \mic.  In fact, these two features, 
by dint of their breadth at the high pressures encountered in brown dwarf atmospheres,
dominate the spectra from $\sim$0.5 \mic to $\sim$1.0 \mic of all dwarfs later than late-Ls.
Since cool atmospheres are partially or totally depleted of the heaviest elements, and these alkali lines
are strong, there are few other significant contributions to the opacity over this octave.   
One of the results is that the combination of absorption by Na-D in the yellow, the low temperatures that 
put the optical in the Wien tail, and the behavior of the red wing of the potassium doublet
all conspire to make it impossible for brown dwarfs to be brown. The color ``brown"
needs some yellow, denied to the object because of the Na-D absorption feature.  The result
is a ``magenta" dwarf, much closer to purple than to brown.

\begin{figure}
   \includegraphics[width=4.0in, angle=-90]{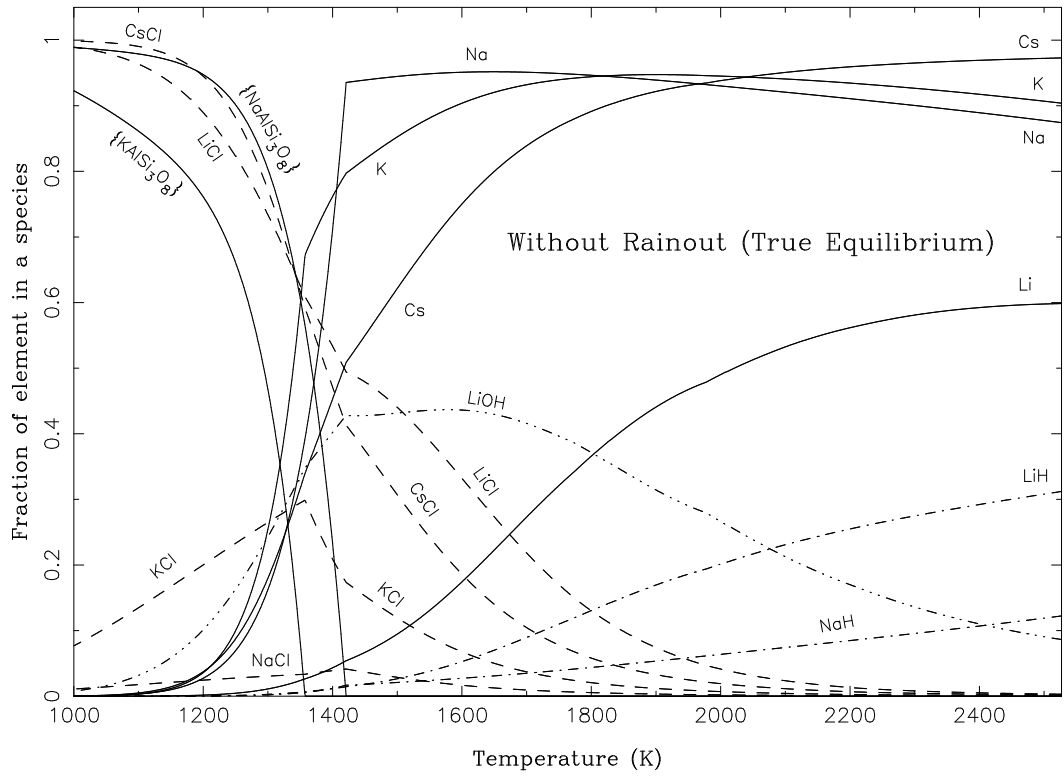}
\caption{The abundances of alkali metal compounds and atoms under the assumption
of chemical equilibrium, without rainout, in a representative brown dwarf atmosphere.
[Taken from Burrows, Marley, \& Sharp 2000]
\label{fig:5}}
\end{figure}

\begin{figure}
   \includegraphics[width=4.0in, angle=-90]{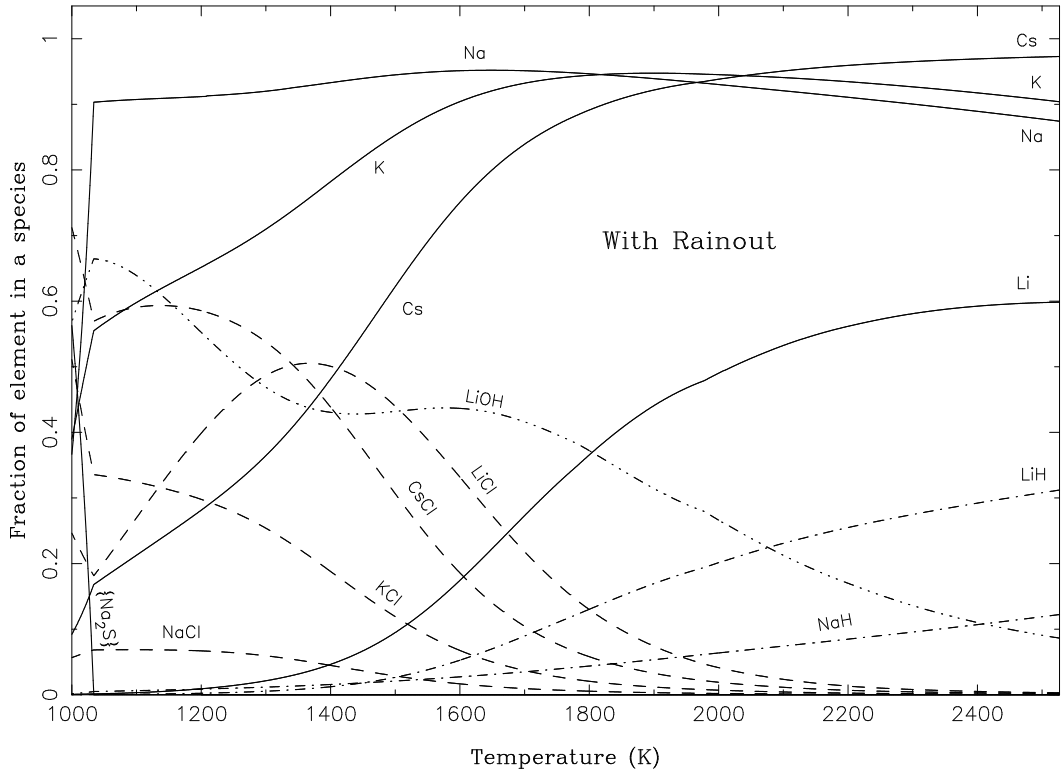}
\caption{The same as Fig. \ref{fig:5}, but with rainout.  Note that the sequestration
of aluminum and silicon in more refractory species deeper in the atmosphere at higher temperatures
undermines the formation of the feldspars of sodium and potassium.  This allows the
atomic form of these alklais to survive to lower temperatures and enhances and extends the predominance
of the resonance lines of atomic sodium and potassium over brown dwarf spectra. It is only 
at still lower temperatures that sodium and potassium (near $\sim$900-1000 K) condense out
into Na$_2$S and KCl and that their effects on brown dwarf spectra from $\sim$0.5 \mic to 
$\sim$1.0 \mic abate. [Taken from Burrows, Marley, \& Sharp 2000]
\label{fig:6}}
\end{figure}

Whither the primacy of these neutral alkali metals?  As discussed above, condensation
of refractory compounds depletes the atmosphere of, for example, Ca, Al, Mg, and Si.
However, Na and K are not as refractory and would condense out at lower temperatures nearer
$\sim$1400 K.  Moreover, they would condense into the feldspars high albite and sanadine,
and such feldspars require silicon and aluminum.  But, since Al and Si elements have already rained out
at higher temperatures and settled, they are not available.  The upshot is that these most abundant of alkalis
persist in their nascent atomic form to even lower temperatures below $\sim$1000 K, at which
point they condense predominantly into Na$_2$S and KCl.  The result is that atomic sodium 
and potassium are in evidence over a broad range of \teffs in the brown dwarf realm.   This,
and their strong spectral influence, combine to determine brown dwarf optical and near-IR 
spectra and colors from \teffs\ of $\sim$1300 K to $\sim$500 K.  Figures \ref{fig:5} and \ref{fig:6}
from Burrows, Marley, \& Sharp (2000) depict examples of the dependence of 
alkali chemistry on atmospheric temperature, without and with rainout.  As a 
comparison between these figures demonstrates, rainout extends the domain 
of importance of atomic Na and K below where eqilibrium chemistry would have depleted them.
In any case, dust plays a central role, however indirect, in the spectra of brown dwarfs.

\section{Conclusions}
\label{conclusions}

The physics and chemistry of dust and clouds have emerged as important components in brown dwarf 
theory.  Dust determines the position of the edge of the hydrogen-burning main sequence, is 
responsible for an entire spectroscopic class (the L dwarfs), and significantly affects 
the abundances and alters the spectra of brown dwarf atmospheres. However, the treatment 
of silicate clouds in theoretical models is still rather primitive.  Therefore, the condensation,
optical properties, grain growth physics, and meteorology of refractories in brown dwarfs
deserves and is getting more attention.  Nevertheless, it is fascinating that something so
prosaic could have so many effects, direct and indirect.  The reader should be aware that 
I have listed here only a few of the most interesting.


\acknowledgements 
This study was supported in part by NASA grants NNG04GL22G and NNX07AG80G.
The author thanks Ivan Hubeny for his many contributions over the years.


\newpage

\end{document}